\documentclass{article}

\usepackage{graphicx} 
\usepackage{geometry}
\usepackage{amsmath}
\usepackage{amssymb}
\usepackage[dvipsnames]{xcolor}
\usepackage{tabularx}
\usepackage{blindtext}
\usepackage{multicol}
\usepackage{authblk}
\usepackage{caption}
\usepackage{makecell}
\usepackage{floatrow}
\usepackage{cite}
\usepackage[label font=bf,labelformat=simple]{subfig}
\title{Local step-flow dynamics in thin film growth with desorption}
\author[1]{Xiaozhi Zhang}
\author[1]{Jeffrey G. Ulbrandt}
\author[2,5]{Peco Myint}
\author[4]{Andrei Fluerasu}
\author[4]{Lutz Wiegart}
\author[4]{Yugang Zhang}
\author[4]{Christie Nelson}
\author[3]{Karl F. Ludwig}
\author[1]{Randall L. Headrick*}
\affil[1]{Department of Physics and Materials Science Program, University of Vermont, Burlington, VT 05405, USA.}
\affil[2]{Division of Materials Science and Engineering, Boston University, Boston, MA 02215, USA.}
\affil[3]{Department of Physics, Boston University, Boston, MA 02215, USA.}
\affil[4]{National Synchrotron Light Source II, Brookhaven National Lab, Upton, NY 11967, USA.}
\affil[5]{X-ray Science Division, Argonne National Laboratory, Lemont, IL 60439, USA.}

\date{}

\begin{document}
\maketitle

\begin{abstract}

Desorption of deposited species  plays a role in determining the evolution of surface morphology during crystal growth when the desorption time constant is short compared to the time to diffuse to a defect site, step edge or kink. However, experiments to directly test the predictions of these effects are lacking. Novel techniques such as \emph{in-situ} coherent X-ray scattering can provide significant new information. Herein we present X-ray Photon Correlation Spectroscopy (XPCS) measurements during diindenoperylene (DIP) vapor  deposition on thermally oxidized silicon surfaces. DIP forms a nearly complete two-dimensional first layer over the range of temperatures studied (40 - 120 $^{\circ}$C), followed by mounded growth during subsequent deposition. Local step flow within mounds was observed, and we find that there was a terrace-length-dependent behavior of the step edge dynamics. This led to unstable growth with rapid roughening ($\beta>0.5$) and deviation from a symmetric error-function-like height profile.  At high temperatures, the grooves between the mounds tend to close up leading to nearly flat polycrystalline films. Numerical analysis based on a 1 + 1 dimensional model suggests that terrace-length dependent  desorption of deposited ad-molecules is an essential cause of the step dynamics, and it influences the morphology evolution.

\end{abstract}

\section*{Introduction}

Step-edge motion has been an area of great interest to scientists and crystal growers for more than seven decades, dating back to the work of Burton, Cabrera and Frank. \cite{Burton1951, Giesen2001} Monitoring nanometer scale surface kinetics during thin-film deposition is one key to understanding the mechanisms of step-edge dynamics. Synchrotron-based  X-ray scattering has proven to be a powerful tool in this area because it is a real-time, contactless, and non-destructive technique that is sensitive to nanometer scale features.\cite{Vlieg, Feidenhansl1989, Robinson1992, Renaud2009} It can be used for \emph{in-situ} measurements where other methods such as scanning probe microscopy are impractical.\cite{Vlieg, fuoss1992}  However, information regarding the dynamics of these processes has not been directly obtainable because conventional X-ray scattering techniques are restricted by spatial averaging \cite{AlsNielsen2011}.  Recently, \emph{in-situ} coherent X-ray scattering probes such as coherent grazing incidence small angle X-ray scattering and near-specular X-ray scattering have proven to be sensitive to local dynamics on surfaces during growth.\cite{Headrick2019, Ju2019 }
 
\begin{figure*}[!ht]
\centering
\includegraphics[width=0.9\textwidth]{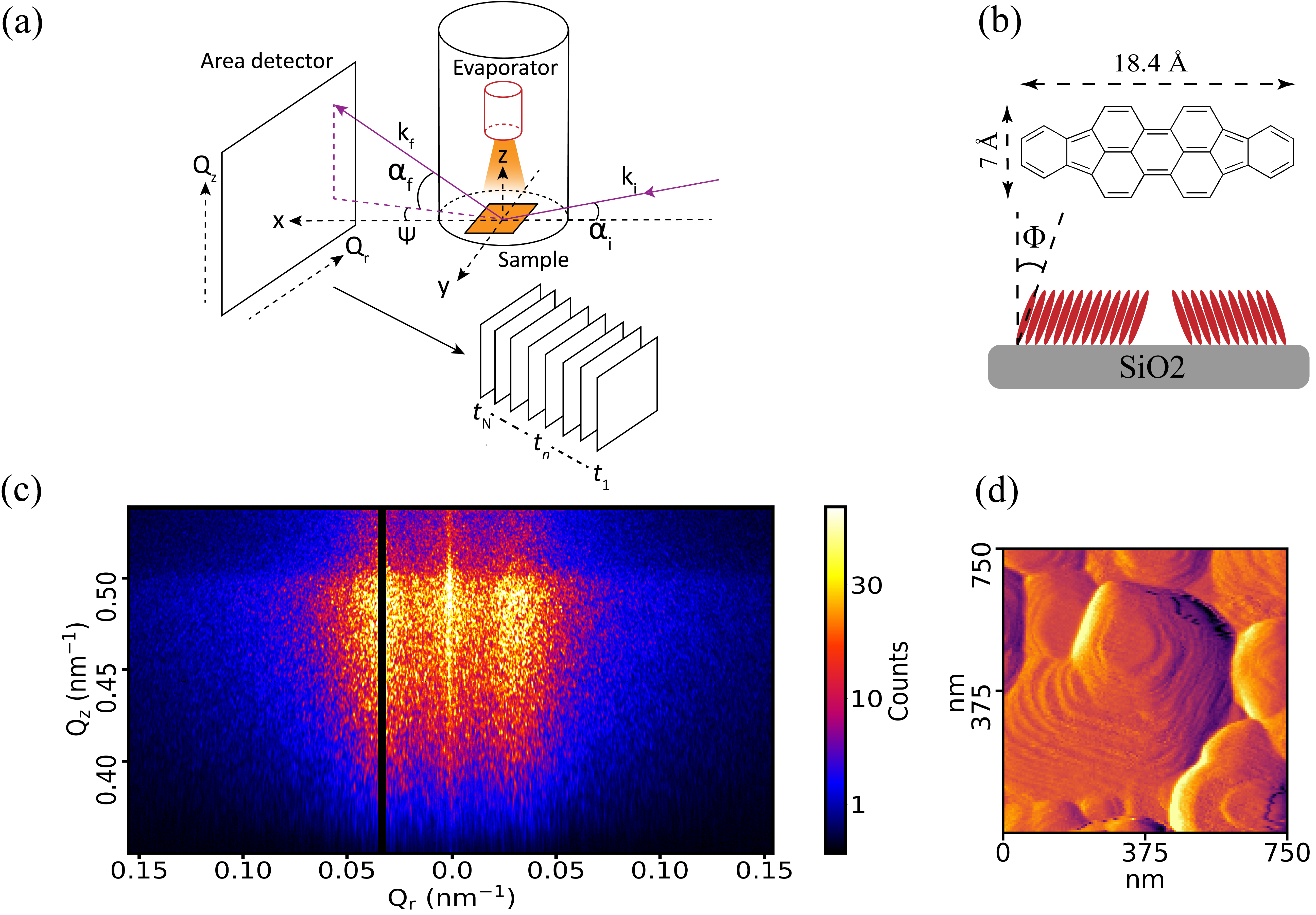}
\caption{(a) Experimental setup of the X-ray scattering measurement. (b) A diagram of the Diindenoperylene molecule and a sketch illustrating the molecule tilt angle $\Phi$ of around 15$^{\circ}$--20$^{\circ}$ \textcolor{black}{with respect to the surface normal in thin films}. (c) Close up view of the scattering pattern taken from the 80 $^{\circ}$C growth at 200 s after the deposition begins. The positions of the broad satellite peaks come from the average mound size within the area illuminated; here it corresponds to $2\pi/Q_{r, peak} \approx$ 208 nm.  Speckle pattern that is unique to the local configuration of mounds and steps is also visible. (d) Post-growth amplitude mode AFM image showing mounds from the film deposited in two layers at 80 and 100  $^{\circ}$C, respectively. Note the molecular-height steps and the flat tops of the mounds.}
\label{Fig1.main}
\end{figure*}

X-ray Photon Correlation Spectroscopy (XPCS) is a spatio-temporal coherent probe that measures the dynamics of nanoscale features from the intensity fluctuation of a speckle pattern produced by X-ray scattering from the sample.\cite{Shpyrko2014} The setup of the XPCS experiment is shown in Figure 1. When coherent light illuminates an object with disorder, the scattered light forms a speckle pattern consisting of irregularly spaced bright and dark spots. An example of a speckle pattern is shown in Figure 1(c). If the disorder is not static, the speckle pattern changes on the same time scale. Thus, by observing fluctuations in the speckle pattern, it is possible to obtain information regarding the dynamics of the corresponding disorder. XPCS is the X-ray probe used to measure the timescale of this speckle dynamics as a function of the X-ray wavevector $\overrightarrow{Q}$. Spatial averaging, which is inevitable during conventional low-coherence X-ray scattering measurements, can be avoided, and the dynamic behavior at a certain length scale corresponding to $2\pi/Q$ can be analyzed.

For a non-equilibrium system, the two-time auto-correlation function can be calculated to probe the system dynamics \cite{Fluerasua, Fluerasu}:
\begin{equation}
G(Q,t_{1},t_{2}) = \frac{\left \langle I'(Q,t_1)I'(Q,t_2) \right \rangle_Q}{\left \langle I'(Q,t_1) \right \rangle_Q\left \langle I'(Q,t_2) \right \rangle_Q}
\end{equation}

$I(Q, t)$ is the X-ray intensity of a particular pixel, which is normalized by $I'(Q,t) = I(Q,t)/\widetilde{I(Q)}$, where $\widetilde{I(Q)}$ is the X-ray intensity averaged over a small range of detector pixels around that particular pixel in order to \textcolor{black} {remove} the speckles. The average $\left \langle \right \rangle$ is taken over a small region of interest (ROI) on the detector \textcolor{black}{at nearly equal $Q$.} 

From the two-time auto-correlation function, correlations between any two different times $t_{1}$ and $t_{2}$ can be obtained. \textcolor{black}{Non-equilibrium dynamics of the system, such as a thermodynamic parameter change, or any sudden changes in the system arising from spontaneous internal or external origins can cause the evolution of the correlations.}

\textcolor{black} {For a system that is close to steady state}, one can assume that the autocorrelation function depends only on the time lag $\Delta t = |t_{1}-t_{2}|$ instead of the specific times  $t_{1}$ and $t_{2}$: 
\begin{equation}
g^{(2)}(Q,\Delta t) = \frac{\left \langle I'(Q,t)I'(Q,t+\Delta t) \right \rangle _t}{(\left \langle I'(Q,t) \right \rangle _t)^2}
\end{equation}

The correlations are then averaged over all available $t$ for each $\Delta t$; the speckle smoothing and averaging over a region of interest are also performed here. The resulting one-time auto-correlation function contains information on the spatial and temporal behavior of the steady state of the system.

%and can also be  expressed as a product of the correlation functions of the electric fields:
%\begin{equation}
%g^{(2)}(Q,\Delta t) = 1+\beta(Q)|F(Q,\Delta t)|^2,
%\end{equation}

%\noindent where $F(Q,\Delta t) = g^{(1)}(Q,\Delta t)/g^{(1)}(Q,0)$, is the normalized intermediate scattering function. $\beta(Q)$ is the optical contrast factor that is directly related to the degree of X-ray beam coherence and experimental setup\cite{pusey1977, Pecora1993}. The intermediate scattering function can typically be expressed as an exponential decay with a relaxation time constant $t_0$: $F(Q,\Delta t)\sim \Gamma_{0}\rm exp\it [-(\frac{\Delta t}{t_{o}})^{n}]$, where $n$ determines the shape of the decay, either stretched ($n<1$) or compressed ($n>1$).

Herein, we present our work on utilizing XPCS to study the growth of  diindenoperylene (DIP) thin films on SiO$_{2}$.  Previous studies have shown that diindenoperylene thin films grown on atomically smooth SiO$_{2}$ substrates have tilt domains with the molecules standing nearly upright with a tilt angle $\Phi$ of approximately 15$^{\circ}$–20$^{\circ}$. Because the base layer of an island can form in any azimuthal orientation, the merging of neighboring mounds is inhibited. Mounds with a  \emph{wedding cake} morphology  form on the surface, and deep valleys appear at the grain boundaries as the films grow.\cite{frank2014} This growth mode has been hypothesized to cause  rapid roughening behavior during deposition due to tilt domains \cite{Duerr}. \textcolor{black} {These properties make DIP thin films a good subject for characterizing the growth process via XPCS. }

\section*{Experiments}

Diindenoperylene (DIP) is a planar perylene derivative with two indeno-groups attached to opposite ends of the perylene core, as shown in Figure 1(b); the chemical formula is C$_{32}$H$_{16}$ and the full chemical name is diindeno[1,2,3-cd:1',2',3'-lm]perylene. The structure of bulk DIP crystals has been studied by Heinrich $et$ $al$ \cite{Heinrich2007}, who identified distinct phases below and above 150 $^{\circ}$C. Below 150 $^{\circ}$C diindenoperylene crystals exhibited a triclinic structure with $a$ = 11.659 $\rm\AA$, $b$ = 13.010 $\rm\AA$, $c$ =  14.966 $\rm\AA$ and $\alpha$ = 98.441$^{\circ}$, $\beta$ = 98.024$^{\circ}$, $\gamma$ = 114.549$^{\circ}$, while above the  transition temperature it transformed to a monoclinic lattice with $a$ = 7.171 $\rm\AA$, $b$ = 8.550 $\rm\AA$, $c$ =  16.798 $\rm\AA$ and $\alpha$ = 90.0$^{\circ}$, $\beta$ = 92.416$^{\circ}$, $\gamma$ = 90.0$^{\circ}$. In thin film growth at 150 $^{\circ}$C by organic molecular beam deposition (OMBD), DIP has been shown to order   with a crystal structure nearly identical to the high temperature monoclinic  phase with  $a$ = 7.09 $\rm\AA$, $b$ = 8.67 $\rm\AA$, $c$ =  16.9 $\rm\AA$ and $\alpha$ = 90.0$^{\circ}$, $\beta$ = 92.2$^{\circ}$, $\gamma$ = 90.0$^{\circ}$.\cite{Duerr2002, Duerr}  While the lattice structure remains close to the high temperature bulk crystal phase, lowering the substrate temperature leads to a change in the molecular orientation of the DIP thin film from a $\sigma$ orientation (standing up with a tilt angle of 15$^{\circ}$--20$^{\circ}$) to a mixture of $\sigma$ orientation and $\lambda$ orientation (lying down) \cite{Kowarik2009, Kowarik, Duerr2002, Duerra}.

DIP was  purchased from ALFA Chemistry and purified by gradient sublimation at the University of Vermont. The SiO$_2$ substrates were cut into 1 $\times$ 1 cm$^2$ pieces and then cleaned with deionized water, acetone, and isopropyl alcohol. The sample was then placed on a heated stage and installed into a custom UHV chamber. The chamber has a downward-facing thermal evaporator with an integrated shutter to control the deposition. The evaporator also had a built-in water-cooled quartz crystal microbalance (QCM) to monitor the deposition rate.  Two 1 cm diameter mica windows were installed on the upstream and downstream ports of the chamber, and the downstream window was offset 0.5 cm upwards to allow X-rays scattered to higher exit angles to reach the detector.

\emph{In-situ} XPCS measurements were carried out at the Coherent Hard X-ray Scattering (CHX) beamline at the National Synchrotron Light Source II, Brookhaven National Laboratory. The chamber was mounted on a diffractometer and a \textcolor{black} {partially} coherent X-ray beam with photon energy of 9.65 keV \textcolor{black} {focused to 10 $\times$ 10 $\mathrm{\mu m^{2}}$} was incident on the substrate surface at 0.4$^{\circ}$ to create a grazing incidence small angle X-ray scattering (GISAXS) pattern. A  Dectris Eiger 4M fast area detector was located 10 m \textcolor{black}{downstream} from the sample in vacuum. \textcolor{black} {The change of the X-ray wavevector $\overrightarrow{Q}$ can be expressed as a function of incident and exit angles \cite{Sinha, Renaud2009} shown in Fig. 1(a):}

\begin{equation}
\overrightarrow{Q} = \overrightarrow{k_{f}} - \overrightarrow{k_{i}} = 
\begin{pmatrix}
Q_{x} \\
Q_{y} \\
Q_{z}
\end{pmatrix}
= \frac{2\pi}{\lambda}
\begin{pmatrix}
\cos(\alpha_{f})\cos(\psi) - \cos(\alpha_{i}) \\
\cos(\alpha_{f})\sin(\psi) \\
\sin(\alpha_{i}) + \sin(\alpha_{f})
\end{pmatrix}
\end{equation}
 \textcolor{black} {In our experimental setup, $Q_{x}$ is small and the horizontal component $Q_{r}$ will be primarily contributed from $Q_{y}$; while $Q_{z}$ is the vertical component. In our work, we will focus on the scattering along and right below the Yoneda wing ($Q_{z}$ $\sim$ 0.4 - 0.5 nm$^{-1}$) because it is particularly sensitive to the surface structure. The ROI map used in the experiment is shown in Supplementary Figure 1.} \textcolor{black}{Based on the experimental setup, in this work our ROI is $\sim$70 $\times$ 70 pixels.}

The deposition was performed by heating the evaporator to approximately 280 $^{\circ}$C to obtain a steady and consistent deposition rate, which was simultaneously monitored by the QCM mounted on the evaporator assembly. In practice, because the temperature of the evaporator is controlled rather than directly controlling the deposition rate, the deposition rate varies by approximately $\pm$10\% from run to run, and it can drift by a similar amount during individual depositions. The deposition rate as a function of growth time measured by the QCM was recorded, as shown in Supplementary Fig. 2 in order to quantitatively consider the variations.  The data collection protocol was as follows: for the \emph{in-situ} XPCS measurements, three samples, each with two layers of film, were grown and  measured. The data collection time for each \emph{in-situ} XPCS scan was 3600 s in total; the shutter that covered the substrate was opened at 40 s and closed again at 3200 s so that the deposition process time was 3160 s.  A separate XPCS data collection scan was performed on each growth at different substrate temperatures, ranging from 40 to 140 $^{\circ}$C with 20 $^{\circ}$C steps. The experimental setup and sample list are presented in Supplementary Table 1.

The thin film samples deposited at CHX were later studied using \emph{ex-situ} X-ray reflectivity (XRR) and grazing incidence X-ray diffraction (GIXD) at the In-situ and Resonant Scattering (ISR) beamline at the National Synchrotron Light Source II, Brookhaven National Laboratory. For these measurements, the samples were mounted on a 6-circle diffractometer at room temperature in air, the incident X-ray energy was 11.34 keV and diffraction patterns were recorded with an Eiger 500K area detector positioned 0.5 m away.

Additional samples were prepared in the same growth chamber in order to study the real-space thin film morphology at various film thicknesses. All samples were characterized using an Asylum MFP-3D Atomic Force Microscope.

\begin{figure*}[!htbp]
\centering
\includegraphics[width=0.45\textwidth]{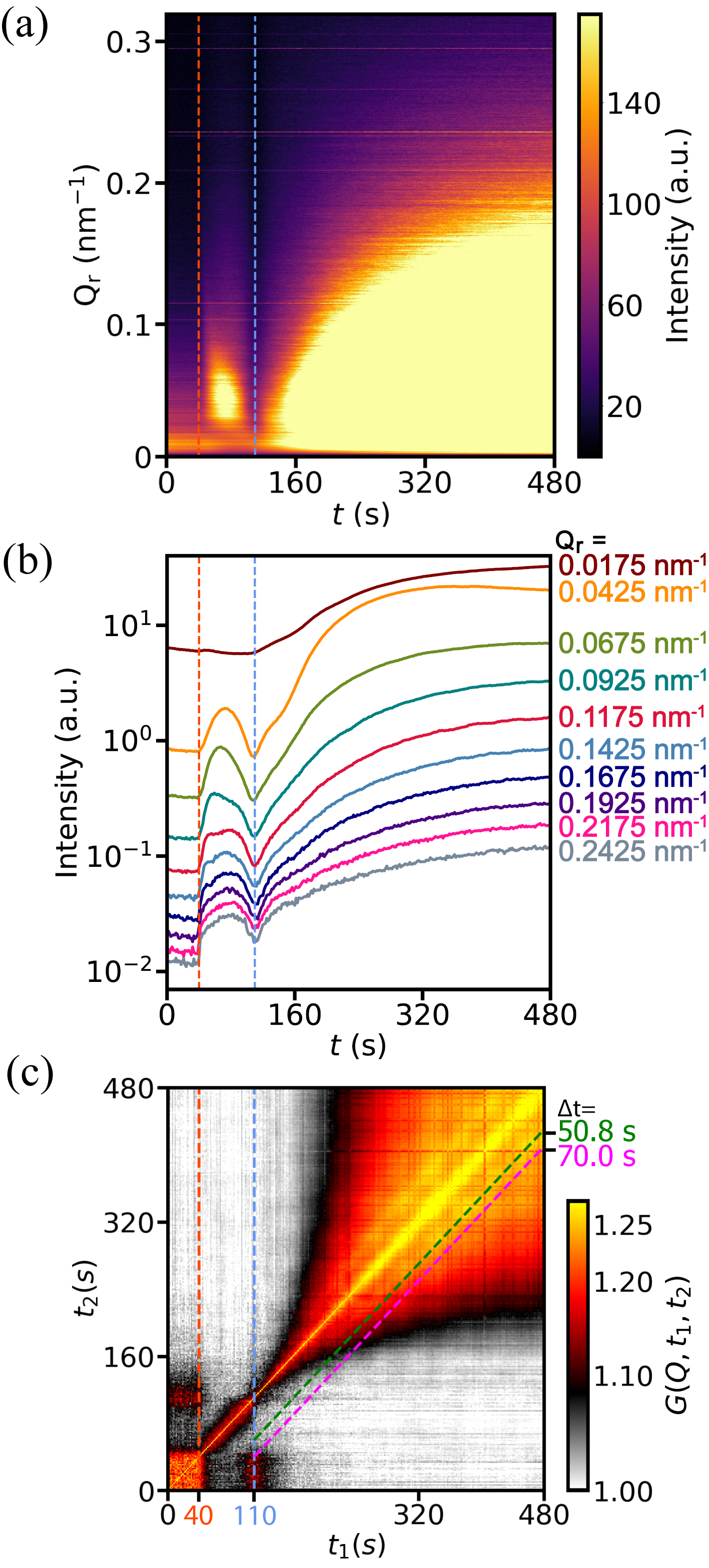}
\caption{ X-ray diffuse scattering profliles and correlations for the first several  layers of growth of Diindenoperylene (DIP) on SiO$_2$, with a substrate temperature of 80  $^{\circ}$C. (a) X-ray intensity of the diffuse scattering profile. (b) Intensity slices from (a). The orange dashed line at 40 s indicates the beginning of the deposition, the blue dashed line at 110 s marks the end of the X-ray intensity oscillation which marks the completion of the first layer growth. (c) Two-time correlation plot ($Q_{r}$ = 0.1425 nm$^{-1}$  ,$Q_{z}$ = 0.4125 nm$^{-1}$) from coherent X-ray analysis. In (c), the magenta dashed line is placed on $t_{2}$-$t_{1}$ = 70 s so it represents the first layer formation time, the green dashed line is overlapped on the streaks in the two-time correlation plot at $T_{corr}=|t_{2}-t_{1}|$ = 50.8 s. Its significance is discussed in the main text.}
\label{Fig2.main}
\end{figure*}

\section*{Results}

\subsection*{In-situ XPCS study}

Figure 2(a, b) shows the X-ray diffuse scattering intensity during the early stages of  the deposition of  DIP on SiO$_2$ at 80 $^{\circ}$C. The diffuse intensity at a specific value of the in-plane scattering vector $Q\rm{_r}$  characterizes the surface roughness at the corresponding length scale $2\pi/Q\rm{_r}$.  After the start of the deposition at $t$ = 40 s, a peak was observed for $Q\rm{_r}  \gtrsim 0.04 \rm{~nm}^{-1}$, which subsided at approximately 110 s. This indicated the completion of the first  DIP monolayer on the SiO$_2$ surface. The growth did not proceed in a layer-by-layer mode since the intensity increased rapidly during the next few deposited monolayers and no intensity oscillations were observed. This intensity evolution is characteristic of a \emph{layer-plus-island} mode, which is similar to the well-known Stranski-Krastanov mode frequently observed in heteroepitaxial growth. Therefore, the DIP ad-molecules  interact differently with the substrate and possibly with the first condensed monolayer  than with the subsequent multilevel film. 

Additional information was obtained from coherent X-ray correlations.  Figure 2(c) shows correlations generated from the same data via the two-time intensity correlation function (TTCF) shown in Eq 1.  The highly correlated regions next to the axes near $t =$ 110 s suggest that the film morphology at that time was correlated with the bare substrate surface before the deposition has started (i.e. $t = $ 0 to 40 s).  This behavior is consistent with the growth of the first DIP layer being close to completion when the next layer nucleates and starts to grow; in other words it  conformally replicates the smooth substrate morphology. The formation of a two-dimensional first layer  was also observed at other growth temperatures (see Supplementary Fig. 3).   After the completion of the first layer, the TTCF  suggests that DIP makes a transition to mounded growth that exhibits \emph{local step-flow}, which for mounded growth is defined as the steps surrounding the apex of each mound flowing outward. This growth mode is associated with streaks in the TTCF parallel to the main diagonal axis,  $\Delta t = |t_1 - t_2| = 0$.   In Fig. 2(c), the streak at $\Delta t$ = 50.8 s is highlighted.  Previous work on polycrystalline C$_{60}$ film growth,\cite{Headrick2019} supported a model in which the streak period $\Delta t = T_{corr}$  is essentially the time interval for steps to advance by one step-spacing.   The streaks originate from  the degree of self-similarity of the step arrays for various time differences, reaching a minimum for time differences that correspond to half-integer monolayer deposition and a maximum for integer monolayer deposition.  We note that this effect is very different from intensity oscillations, as the oscillations are in the correlations rather than in the scattering intensity, and it is the  \emph{time difference} that is important because there is no specific time when an entire layer reaches completion during mounded growth.

The first layer formation time, which we denote as   $T_{\rm{1st}}$ is also highlighted  in Fig. 2(c) and  it is observed to be significantly longer than $T_{corr}$ (70.0 s vs. 50.8 s).  A plausible explanation for this difference is that because SiO$_{2}$ is an inert substrate the sticking coefficient of DIP is less than one, so it requires a longer time to form a complete first layer.   Table 1 shows the effective sticking coefficients for three different growth temperatures derived from the ratio of the first layer formation rate $1/T_{\rm{1st}}$  to the expected rate $1/T_{\rm{1st,est}}$  from the calibrated QCM data.  The results show that the ratio is less than one and it decreases from $\approx$ 0.8 to 0.7 as the growth temperature is increased from 40 to 120 $^\circ$C.  \textcolor{black}{Note that  the ratio $T_{\rm{1st,est}}/T_{\rm{1st}}$ is not strictly equal to the sticking coefficient on the substrate since a significant fraction of the deposited molecules land on top of recently formed monolayer islands, and they can be subsequently incorporated into the first layer at the island boundary.} A  similar estimate for the bulk film sticking coefficient (Table 1, last column)  was derived by dividing the final film thickness by the integrated QCM rate; \textcolor{black} {see Supplementary Tables 1 and 2 for detailed information of calculating the bulk film sticking coefficient}. The results revealed that the bulk film sticking coefficient also decreased as the growth temperature increased.    However, the desorption effect was not as strong  as that during the first layer deposition.

\begin{table*}[!htbp]
\centering
\newcolumntype{Y}{>{\centering\arraybackslash}X}
\begin{tabularx}{1.0\textwidth}{YYYYYY}
\hline \hline \\
Temperature ($^{\circ}$C)     & Estimated first monolayer deposition rate, $R_{d,1st}$ (nm/min)$^{a}$  & Estimated first monolayer formation time, $T_{1st, est}$ (s) $^{b}$  & Experimental first layer formation time, $T_{1st}$ ~~(s)  &  \textcolor{black} {Ratio of $T_{1st,est}$ and $T_{1st}$}  & Bulk film sticking coefficient$^{c}$  \\ \\ \hline \\
40    & 1.865 & 53.5  &  65   &  0.82 & 0.99     \\ \\
80    & 1.696 & 58.8  &  70   &  0.84 & 0.93     \\ \\
120  & 1.888 & 52.8  &  77   &  0.69 & 0.78    \\ \\  \hline \hline
\multicolumn{6}{>{\hsize=7\hsize}X}{ 
$^{(a)}$Derived from QCM data. 
$^{(b)}$Calculated from $d/R_{d, 1st}$ neglecting desorption, where $d$=1.663 nm is the layer spacing.  
$^{(c)}$Derived from QCM data and the final film thickness measured by profilometry, see Supplementary Tables 1 and 2 for details.}\\ 
\hline
\end{tabularx}
\caption{Bulk sticking coefficient and first layer formation rates for DIP films deposited on SiO$_{2}.$}
\label{table:1}
\end{table*}

To discuss the step motion  in more detail, we  distinguish between the correlation period $T_{corr}$ that was introduced above, and the mean monolayer formation time estimated from the film thickness, which we denote as simply $T$.     Because there is some unintended variation in the deposition rate, we averaged the monolayer formation time over selected time intervals where the deposition rate is most stable and we used the QCM data  to correct $T$ for each specific time interval.  For example, in the time interval used in the analysis of Fig 2(c), 320-640 s, the monolayer completion time calculated from the combined QCM and profilometry data is  $T = 54.9 \pm$ 0.3 s.  Growth rate data can be found in Supplementary Figure 2.  

The analysis above describes three characteristic times: $T_{1st}$, $T_{corr}$ and $T$. While the first of these is clearly different because it pertains to the first deposited monolayer, the distinction between $T$ = 54.9 s and  $T_{corr}$ =  50.8 s is more subtle.   We have found that  the correlation period is a function of $Q_r$, such that, while $T$ is a measure of the overall monolayer formation time. $T_{corr}$ in Fig. 2(c) is interpreted as a measure of the local growth rate for features of the size corresponding to $2 \pi / Q_r$, approximately 44 nm.  We equate these features \textcolor{black}{($\sim$44 nm)} with the step spacings. Figure 3(a) shows the complete two-time autocorrelation plot of the same deposition at 80 $^{\circ}$C calculated from Eq. 1. Two different time intervals are indicated by the dashed boxes,  320-640 s  and  800-1600 s.  In Figure 3(b), the period $T_{corr}$  of the oscillation is plotted as a function of 2$\pi /Q_r$ (Figure 3).   The correlation period for larger length scales is close to the nominal monolayer deposition time $T$, whereas smaller length scales have a shorter period.  We propose that this behavior is related to terrace-length-dependent desorption. A fitting function is  applied  in  Fig. 3(b), which is pertinent to growth with desorption, as discussed in the Computational Modeling subsection. The  difference between $T$(320-640 s) = 54.9 $ \pm$ 0.3 s and $T$(800-1600 s) =  67.7 $\pm$ 0.4 s is due to a drifting of the deposition rate, although the variation within each time interval is much smaller, as reflected in the uncertainties.

\begin{figure*}[!htbp]
\centering
\includegraphics[width=0.9\textwidth]{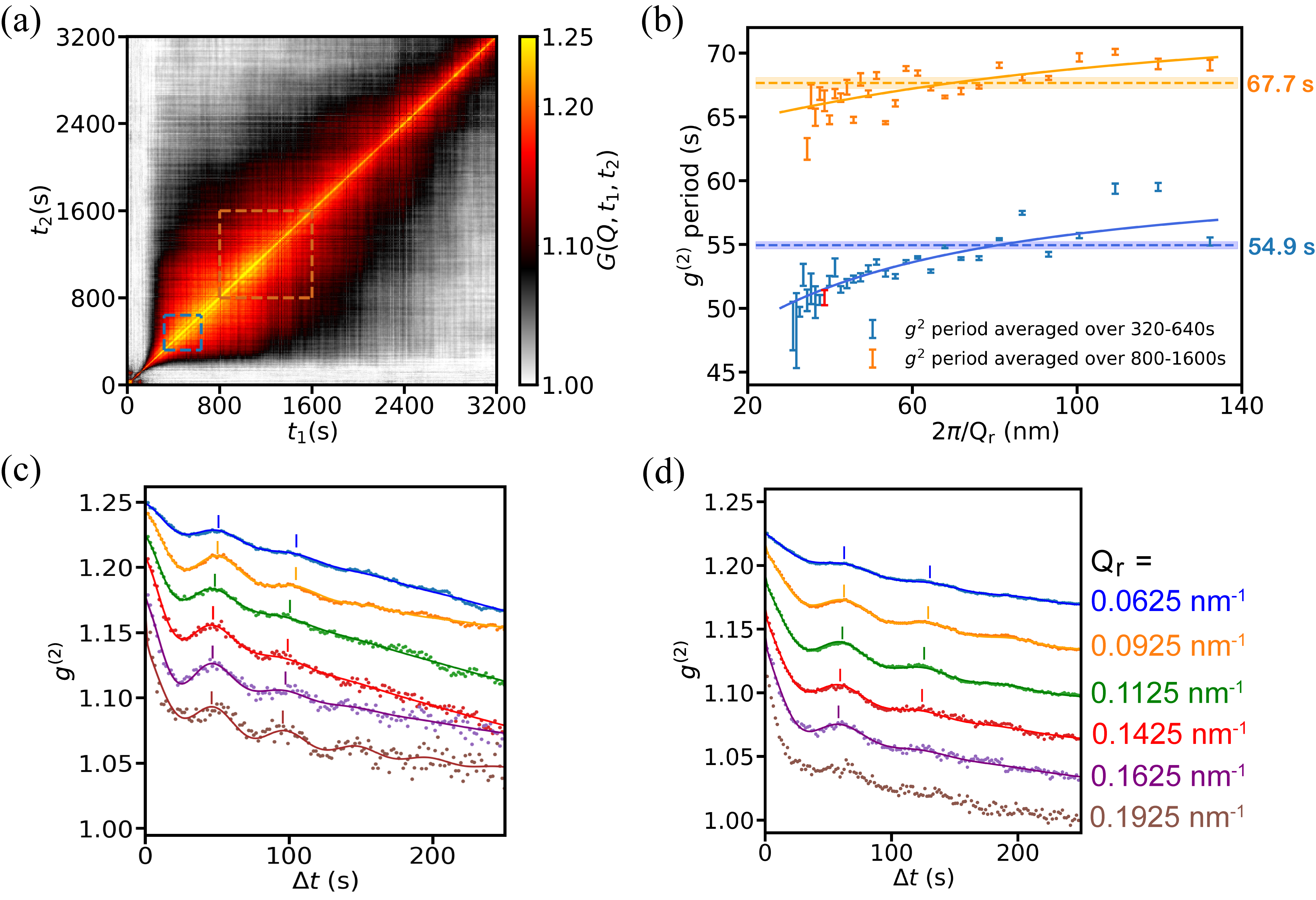}
\caption{XPCS analysis of the DIP deposition on SiO$_2$, the data used here are from the same deposition as Figure 2. (a) The complete deposition at 80 $^{\circ}$C, shutter opens at 40 s and closes at 3200 s ($Q_{r}$ = 0.1425 nm$^{-1}$; $Q_{z}$ =  0.4125 nm$^{-1}$). (b) Oscillation period of autocorrelation plotted as a function of 2$\rm\pi$/$Q_{r}$, the data point for length scale 2$\rm\pi$/$Q_{r}$ =  44.1 nm and period $T_{corr}$ = 50.8 s is highlighted in red. The dashed lines mark the nominal monolayer formation times $T$ calculated from the estimated deposition rates, while the shaded areas show their standard deviation of the mean. The data is fitted to obtain the terrace-length-dependent sticking parameters  $B$ = 0.27 and $L_{d}$ = 16.7 nm for 320 - 640 s, and $B$ = 0.14 and $L_{d}$ = 41.9 nm for 800 - 1600 s. \textcolor{black} {See the Computation Modeling subsection for details.}  (c)  Autocorrelations averaged over 320 s to 640 s as a function of $Q_{r}$ and $Q_{z}$ were fixed at 0.4125 nm$^{-1}$. (d) Autocorrelations averaged over 800 s to 1600 s. \textcolor{black} {Time intervals in (c) and (d) are chosen due to the stability of the deposition rate in those intervals.} \textcolor{black} {Curves in (c) and (d) are shifted for better visibility.}}
\label{Fig3.main}
\end{figure*}

The $Q_{r}$ dependence  can be observed directly from the one-time correlation curves in Figure 3(c) and (d) for 320-640 s and 800-1600 s, respectively. Markers have been added to highlight  the period of a full oscillation. Oscillations existed through the entire growth process, although their amplitudes decayed slowly as the deposition proceeded. No corresponding oscillations were observed in the diffuse scattering intensity (Supplementary Figure 8), confirming that the correlation oscillations do not originate from a layer-by-layer growth process.

\begin{figure*}[!htbp]
\centering
\includegraphics[width=0.8\textwidth]{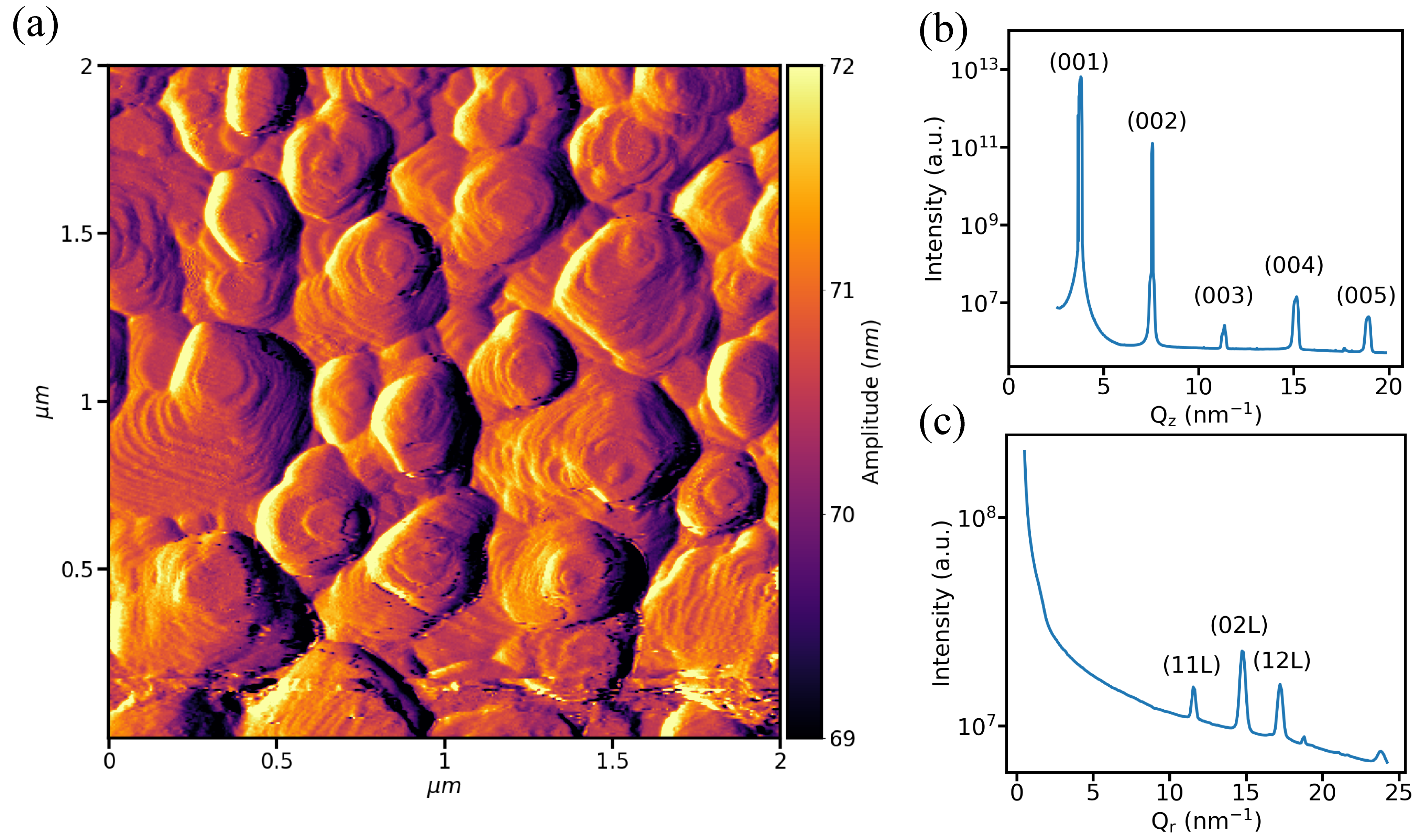}
\caption{ Post-deposition characterization of DIP film on SiO$_2$. (a) \textcolor{black} {AFM image of a thin film after 2 depositions, 80 and 100 $^{\circ}$C for the first and second layers respectively.} The total film thickness is 166.99 nm, which is about 103 layers. (b) X-Ray Reflectivity characterization of the same sample, the film layer spacing is found to be 1.662 nm. (c) Grazing incidence X-ray diffraction of the same sample.}
\label{Fig4.main}
\end{figure*}

The data in Fig. 3 (c, d) were modeled with an empirical intermediate scattering function of the form:

\begin{equation}
g^{(1)}(Q,\Delta t) = \Gamma_{0} \exp \left(-\left[\frac{\Delta t}{\tau_{0}}\right]^{n}\right) + \Gamma_{1} \exp \left(i\frac{2\pi\Delta t}{T_{corr}} - \frac{\Delta t}{\tau_{1}}\right)
\end{equation}

\noindent
\textcolor{black} {The first term $\Gamma_{0}\exp\it (-[\frac{\Delta t}{\tau_{0}}]^{n})$ resembles an exponential decay that has been widely observed in many other systems with a relaxation time constant $\tau_{0}$; where $n$ determines the shape of the decay, either stretched ($n<1$) or compressed ($n>1$). The second term represents the smaller phase-advancing term where $T_{corr}$ is the ($Q_r$-dependent) oscillation period and $\tau_1$ is its own relaxation time constant; this form is close to the heterodyne form used to describe capillary waves on liquid surfaces \cite{Gutt}. Eq. 4 can be inserted into:
\begin{equation}
g^{(2)}(Q,\Delta t) = 1+\beta(Q)|F(Q,\Delta t)|^2,
\end{equation}
\noindent
to obtain a model for the one-time correlation function. Here, $F(Q,\Delta t) = g^{(1)}(Q,\Delta t)/g^{(1)}(Q,0)$, is the normalized intermediate scattering function. $\beta(Q)$ is the optical contrast factor that is directly related to the degree of X-ray beam coherence and experimental setup\cite{pusey1977, Pecora1993}.} 
The fitting parameters for the 80 $^{\circ}$C results shown in Fig. 3 are listed in Supplementary Table 4. This form has previously been used  to describe  XPCS data for  C$_{60}$ growth on graphene/SiO$_2$, which also has a mounded structure with steps. \cite{Headrick2019}. \textcolor{black} {Deposition 40 and 60 $^{\circ}$C exhibited similar correlations with much weaker oscillations, indicating that local step flow occurs over a range of growth temperatures. See Supplementary Fig. 4 for details.}

Figure 4 shows post-deposition characterization of the sample after sequential 80 and 100 $^{\circ}$C depositions.    Figure 4(a) shows an AFM image of the surface plotted in amplitude mode  to highlight the steps and terraces.  The  mounds have an overall size that ranges from about 250 to 550 nm in diameter, while the step spacings range to well below 100 nm.  The values of 2$\rm\pi /Q_r$ in Figure 3(b) span  similar values,  consistent with  the coherent oscillations being associated with the step motion during the growth.   Figure 4(b) shows an X-Ray Reflectivity (XRR) scan of the same sample; only (00$l$) reflections are observed and this suggests the film is highly oriented. The layer spacing can be calculated from the (00$l$)  spacing and is found to be 1.662 nm.  In-plane X-ray diffraction on the sample shown in Fig 4(c) confirms that the film is well ordered. It is notable that we do not observe the $\lambda$ (lying down) orientation for any growth temperature. The AFM height scan and XRR scans for other growth temperatures are in Supplementary Figures 5, 6 and 7. The complete set of lattice constants extracted from the combined in-plane and out-of-plane scans is presented in Supplementary Table 3.

\subsection*{AFM study of the morphology evolution versus film thickness}

\begin{figure*}[!htbp]
\centering
\includegraphics[width=0.7\textwidth]{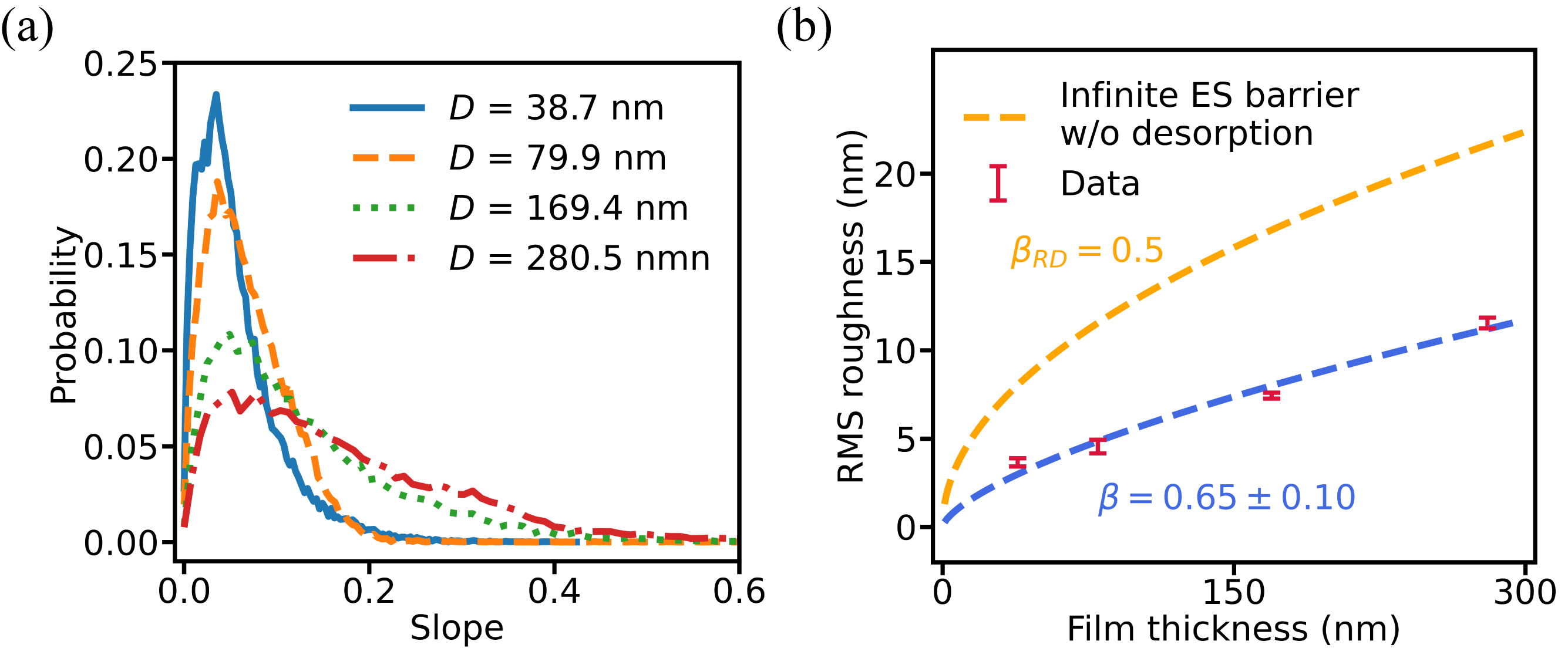}
\caption{Results from post deposition AFM scans of samples with different deposition times. (a)  Probability distribution of slopes \textcolor{black} {calculated through local plane fitting} \textcolor{black} {at different film thickness.} (b) RMS roughness versus film thickness. Power laws fit and exponents are added for comparison. }
\label{Fig5.main}
\end{figure*}

To further study the mound shape evolution during the DIP thin film deposition, four additional samples  with different deposition times were prepared at our home lab.  Figure 5 shows the results of the AFM measurements of the four samples. Thicker films exhibited progressively higher slope features  [Fig. 5(a)],  and the RMS roughness  also continuously increased [Fig. 5(b)]. \textcolor{black} {The AFM images of these samples and slope calculating method are shown in Supplementary Fig. 9.}     The mound profiles tended to be flat-topped with high-slope regions in the valleys between adjacent mounds.

Mounded growth is primarily driven by step-edge barriers that prevent molecules from hopping down to lower levels.\cite{Ehrlich1966, Schwoebel1966, Schwoebel1969,Politi}  For small critical radii below a threshold value, the mound shape becomes unstable and continues to evolve during growth,  resulting in \emph{wedding cake} shaped mounds.\cite{Krug2002}  In the case of an infinite step-edge barrier, the asymptotic shape of the mound expressed  in terms of the layer coverage profile $\theta_{n}$ has a closed form\cite{Cohen1989}: 

\begin{equation}
\begin{gathered}
\theta_{n}(t) = 1 - C[1 + \text{erf}(s/\sqrt{2})]\\s = (n - Ft)/\sqrt{Ft}
\end{gathered}\label{eq:erf_profile}
\end{equation}

\noindent
where $F =  R_d / d$ is the deposition flux \textcolor{black} {with $R_d$ being the deposition rate and $d$ being the layer spacing.} $t$ is the total growth time,  $n$ stands for the $n$th layer, and $C$ is the constant of integration.  Infinite step-edge barriers constrain the molecules to remain at the level where they land on the surface. Even if they diffuse laterally, the height profile is characteristic of  random deposition (RD) where the roughness increases as $\sigma = d(Ft)^{\beta_{RD}}$ with a growth exponent  $\beta_{RD} = 1/2$.  

We found that the magnitude of the roughness was less than that predicted by random deposition, as shown in Fig. 5(b). This implies that the step-edge barriers are finite, which seems reasonable given that \textcolor{black} {in a growth} the activation barrier for interlayer transport is likely to be less than that for desorption.  Another observation is that a  roughening exponent $\beta$ $\approx$ 0.65 is obtained from the data.   This behavior is reminiscent of previous results by D\"{u}rr et al.  who reported an unusually rapid increase in vertical roughness during DIP growth at 145 $^{\circ}$C with $\beta\approx$ 0.75. \cite{Duerr}  A large value of $\beta$ implies that the roughness will eventually overtake the roughness predicted by random deposition.  Roughness that overtakes the RD limit is inconsistent with a model that involves only step-edge diffusion and relaxation by interlayer transport, which promotes smoothening rather than roughening. Below, we discuss whether terrace-length dependent desorption can produce an instability that induces rapid roughening. 

\subsection*{Computational modeling}

\begin{figure*}[!htbp]
\centering
\includegraphics[width=0.9\textwidth]{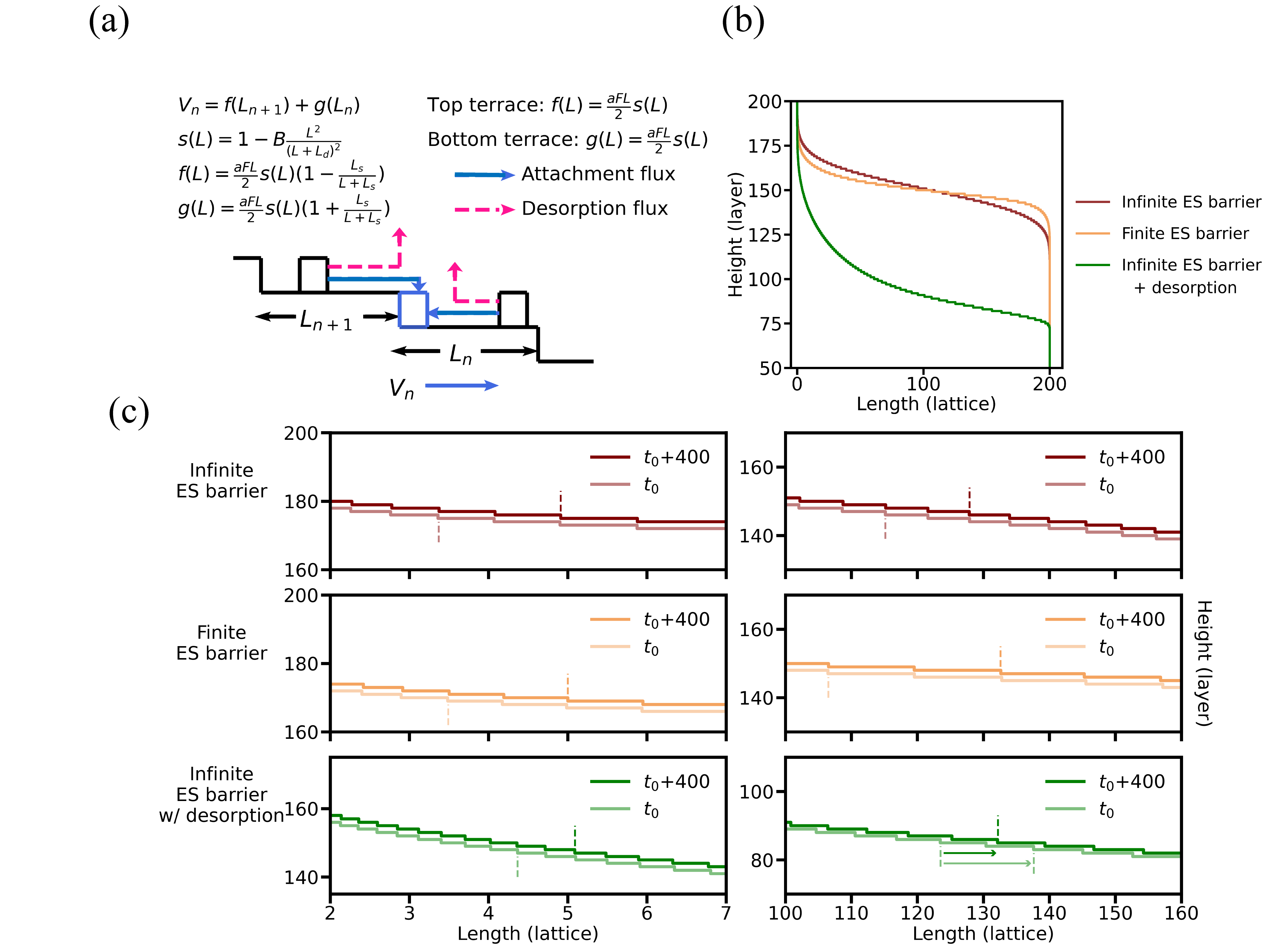}
\caption{ (a) Schematic diagram and equations for Local Step Flow. (b) Comparison of growth models with 150  deposited layers generated on a 200 lattice unit base. The parameters are $F$ = 0.005 and $L_{c}$ = 0. For the \emph{Infinite ES barrier} case $L_{s}$ = $\infty$, while for a \emph{Finite ES barrier} it is 8 lattice units. For the case \emph{Infinite ES barrier + desorption}, the terrace-length-dependent desorption effect is introduced by setting $L_{d}$ = 2 and $B$ = 0.5.  The average film thickness is less than 150 for this case due to desorption. (c) Comparison of step propagation. For each model, two corresponding profiles are  plotted with a separation of 400 time steps, which equals two monolayers in the absence of desorption. The 3 plots on the left highlight step edges with small terrace lengths ($<$ 2 lattice units), while  the 3 on the right focus on longer step edges ($>$5 lattice units). The annotations highlight that the steps advance by approximately two terrace lengths in all cases, except for the longer terraces when desorption is active (\textcolor{black} {$\sim$1.2 terrace length in the} lower right panel).}
\label{Fig6.main}
\end{figure*}

In our previous work on C$_{60}$ mounded growth,\cite{Headrick2019}  we developed a Local Step Flow (LSF) model, which applies the 1 + 1 dimensional \emph{Zeno model}  introduced by Politi and Villain \cite{Politi} to describe the step motion. In the Local Step Flow model, the XPCS correlation oscillation period is interpreted as the time for steps to advance one terrace length; that is,  each step edge advances to the horizontal position of the step edge ahead of it while the mound height increases by one monolayer.  Thus,  $T_{corr} = L_{n}/v_{n}$, where $L_{n}$ and $v_{n}$ are the length and velocity of the $n$th terrace and step, respectively. Generally, the mounds approach a quasi-stationary shape so that all steps take the same amount of time to traverse their own terrace length. As a result, there is no length-scale dependence of $T_{corr}$. In contrast, there is a clear $Q_r$ dependence in our data for DIP growth, suggesting that an essential feature of the step edge dynamics is missing from the model.

In LSF, the strength of the Ehrlich-Schwoebel (ES) step-edge barrier  \cite{Ehrlich1966, Schwoebel1966, Schwoebel1969} is incorporated through a single parameter, the Schwoebel length $L_{s}$. If the terrace length $L$ is much smaller than $L_{s}$, most of the material landing on this terrace goes to its upper edge and if $L$ $\gg$ $L_{s}$  is used, about half of the material goes to the upper edge and the other half to the lower edge. Then the velocity $v_{n}$ of the $n$th step edge is determined by its own terrace length $L_{n}$ and its upper terrace length $L_{n+1}$. Equation \ref{velocity_eqn} shows the equation of motion of the $n$th step in terms of the local attachment velocity contributions from the upper and lower terraces, $f(L_{n+1})$ and  $g(L_n)$, respectively. 

\begin{equation}   v_{n} = f(L_{n+1}) + g(L_{n}) \label{velocity_eqn} \end{equation}

The complete set of equations is shown in Fig. 6(a). Note that the attachment velocity contribution from the upper terrace of a given step $f(L)$ goes to zero and $g(L)$ goes to $aFL$ in the limit $L\rightarrow 0$, in accordance with the rule stated above, and in the limit $L\rightarrow \infty$ they both approach $aFL/2$.  The particle flux per lattice site is $F$, where $FL$ represents the total deposition flux on a terrace of length $L$. \textcolor{black}{Another key parameter of the LSF is the critical radius $L_{c}$: once the radius of the current topmost layer reaches $L_{c}$, a new top layer will start to nucleate.}

% Hidden because they repeat the formulas in the figure.
%\begin{flalign*}  &\textrm{Step velocity:} &v_{n} &= f(L_{n+1}) + g(L_{n})&&  \end{flalign*}
%\begin{flalign*}         &\textrm{Downward flux:}  &f(L) &= \frac{aFL}{2}s(L)\left(1-\frac{L_s}{L+L_{s}}\right)&& \end{flalign*} 
%\begin{flalign}        &\textrm{Upward flux:}~~~~~~~~       &g(L) &= \frac{aFL}{2}s(L)\left(1+\frac{L_s}{L+L_{s}}\right)&&  \end{flalign}
%\begin{flalign*}         &\textrm{Top step:}~~~~  &f(L) &= \frac{aFL}{2}s(L) && \end{flalign*} 
%\begin{flalign*}        &\textrm{Bottom step:}   &g(L) &= \frac{aFL}{2}s(L) &&  \end{flalign*}

It has been argued that even weak desorption can have a pronounced effect on thin film morphology \cite{Villain1991, Lo2002, Smilauer1999}. Here we studied the effect of desorption by modifying the LSF step-edge arrival factors  using a terrace-length-dependent sticking function $s(L)$ \cite{Smilauer1999}:

\begin{equation}
s(L) = 1 - B \frac{L^{2}}{(L+L_{d})^2}
\label{eq:sticking_function_eqn}
\end{equation}

\noindent
where $B$ is a number between 0 and 1, which characterizes the overall strength of the desorption, and $L_{d}$ is the desorption length. This function modifies the  \emph{local attachment flux} such that for $B>0$,  it is smaller than the \emph{local deposition flux} $FL$.

Figure 6 shows several examples  of mound shapes with and without desorption.  The lattice unit $a$ is given a value of unity and the deposition flux $F$ is 1/200, so that in the absence of  desorption the monolayer growth time is exactly 200 time steps. In Figure 6(b), a mound consisting of 150 complete layers of material with infinite ES barrier ($L_{s} =  \infty$), critical radius of zero ($L_{c}$ = 0), and no desorption ($B$ = 0) produces an error function shaped surface profile, corresponding to the closed form expression of Equation \ref{eq:erf_profile}.      For comparison, Figure 6(b) also  shows: (i)  the effect of a finite ES barrier with $L_{s}$ = 8 lattice units and no desorption ($B$ = 0), and (ii)   an infinite ES barrier ($L_{s}$ = $\infty$) with strong desorption ($B$ = 0.5, $L_{d}$ = 2). The profiles illustrate that a finite ES barrier (case i) makes the shape flatter but still symmetric to the middle point of the profile. In contrast,   terrace-length-dependent desorption strongly flattens the mound near the lower edge, while the top of the mound becomes peaked, leading to an asymmetry of the mound profile.   It is notable that with zero critical radius of the next layer can nucleate as soon as the topmost layer starts growing, which leads to a pronounced asperity at the apex of the mound. The growth rate is highest near the peak precisely because the terrace lengths are shortest where the slope is the highest.

\begin{figure*}[!htbp]
\centering
\includegraphics[width=0.65\textwidth]{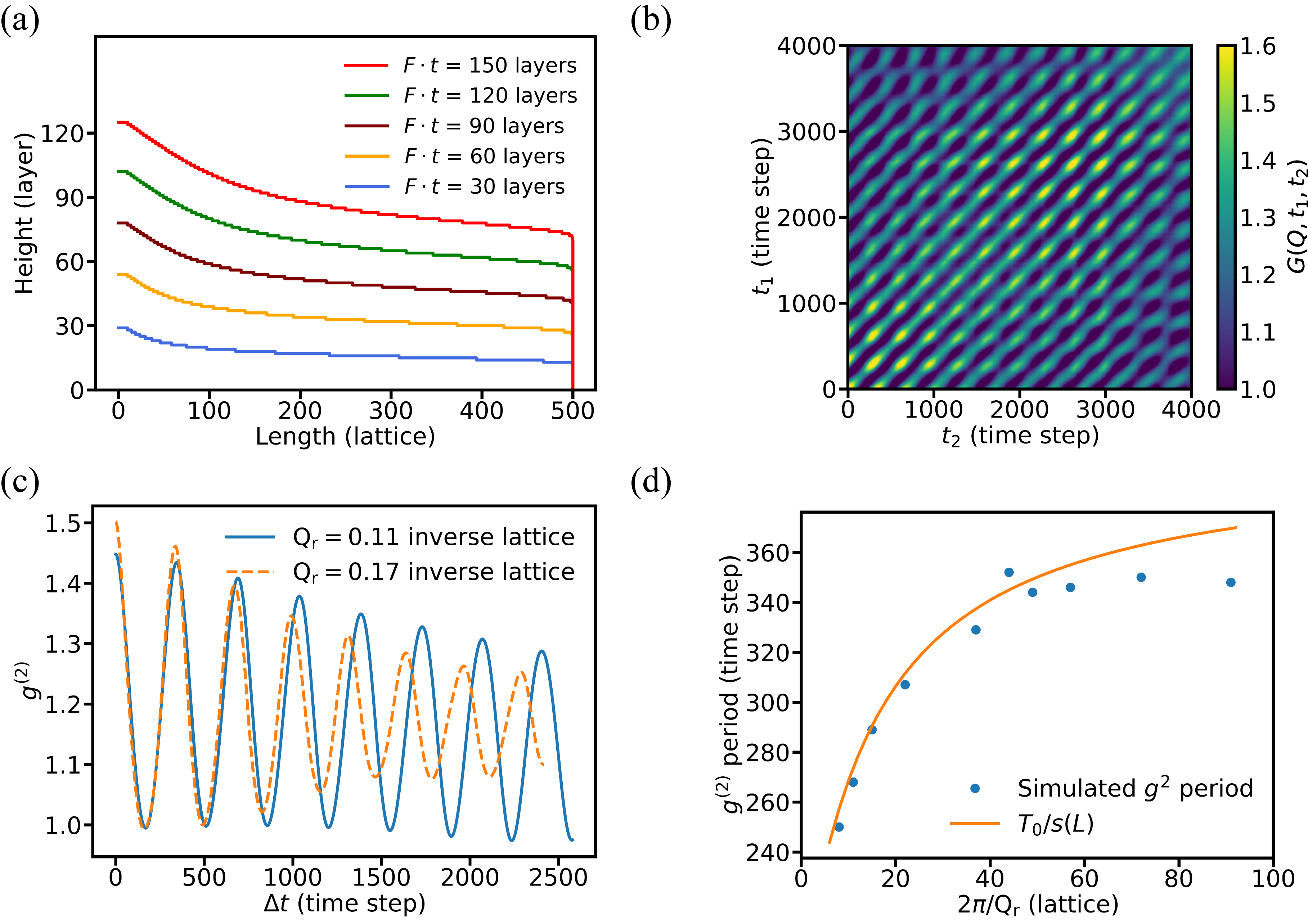}
\caption{Numerical results for the LSF model. (a) Mound profiles for several growth times with parameters: $F$ = 0.005, $L_s$ = 50, $L_c$ = 10, $L_d$ = 4, and $B$ = 0.5  (b) Two-time autocorrelations calculated at 0.11 inverse lattice units, which corresponds to a length of 57 lattice units.  (c) One-time autocorrelations at 0.11 inverse lattice units and 0.17 inverse lattice units (37 lattice units). (d) The $Q_{r}$  dependent  oscillation period obtained from numerical results. The line shows the \textcolor{black} {estimated} monolayer formation time calculated from $T_{0}/s(L)$, \textcolor{black} {where $T_{0}$ = 200 time steps is the monolayer formation time without desorption}. }
\label{Fig7.main}
\end{figure*}

Fig 6(c)  confirms  that the step edges advance by two terrace spacings for the deposition of two full monolayers when desorption is not present.   When desorption is present (lower two plots), the longer terraces that satisfy the condition $L \gg L_s $ lag.  The marks in the lower right panel show that the steps associated with longer terraces advance a little more than one terrace length for every two deposited monolayers. This agrees with  Eq. \ref{eq:sticking_function_eqn}, which was used to generate the mounds.

% So the step edge dynamics of each growth models need to be examined. In Figure 6(c), two mounds with 400 time steps difference of growth time are generated under each 3 conditions: infinite step-edge barrier, finite step-edge barrier and  infinite step-edge barrier plus desorption. The 3 plots above are zoomed in on where the step edges are close ($<$ 3 lattice unit) and the 3 below are focused on step edges are far away ($>$ 10 lattice units). If there's no desorption, 200 time steps should be the monolayer formation time since $F$ = 0.005, the results show if only the step-edge barrier is present, whether it is an infinite barrier or a finite barrier, all the step edges from XXX dt = 400 time steps XXX, while with terrace dependent desorption the more separated step edges will fall behind while the closer ones are still almost lined up.  XXX should advance two terrace lengths XXX

Figure 7 shows autocorrelation results for a simulated mound with desorption. Figure 7(b) is calculated using Eq. 1 for 2$\rm\pi$/$Q_{r}$ $\approx$ 57 lattice units; this range is close to the largest terrace length from the simulated mound shown in Figure 7(a). Parallel diagonal streaks that correspond to the time interval of step edges advancing by their own terrace length can be seen in the simulated two-time autocorrelation plot. An average over aging time $t_{age} = (t_{1} + t_{2})/2$ is performed to obtain the one-time autocorrelation results, which is plotted in Figure 7(c) as the blue curve; the oscillation period of this one-time autocorrelation is found to be 346 time steps. Another one-time autocorrelation is calculated over a range of 2$\rm\pi$/$Q_{r}$ $\sim$ 37 lattice units and show a smaller period  of 329 time steps. Both periods are significantly larger than the nominal monolayer growth time without desorption (200 time steps).

A series of  simulated autocorrelation oscillation periods were plotted as a function of 2$\rm\pi$/$Q_{r}$ in Figure 7(d).  \textcolor{black} {The line in Figure 7(d) shows the estimated monolayer formation time calculated from $T_{0}/s(L)$. $T_{0}$ is the desorption free monolayer formation time, which should equal to 1/$F$ = 200 time steps. $s(L)$ is the sticking coefficient from Equation \ref{eq:sticking_function_eqn} using the values of $B$ and $L_d$ that were input into the simulation.}  Agreement with the computational results is evident, confirming that the desorption parameters can be estimated from the correlation data. We note that $B$ is 2-3 times larger than the values used to model the experimental results  in Figure 3(b) in order to highlight the effect of desorption in the simulation.  

\begin{figure*}[!htbp]
\centering
\includegraphics[width=0.7\textwidth]{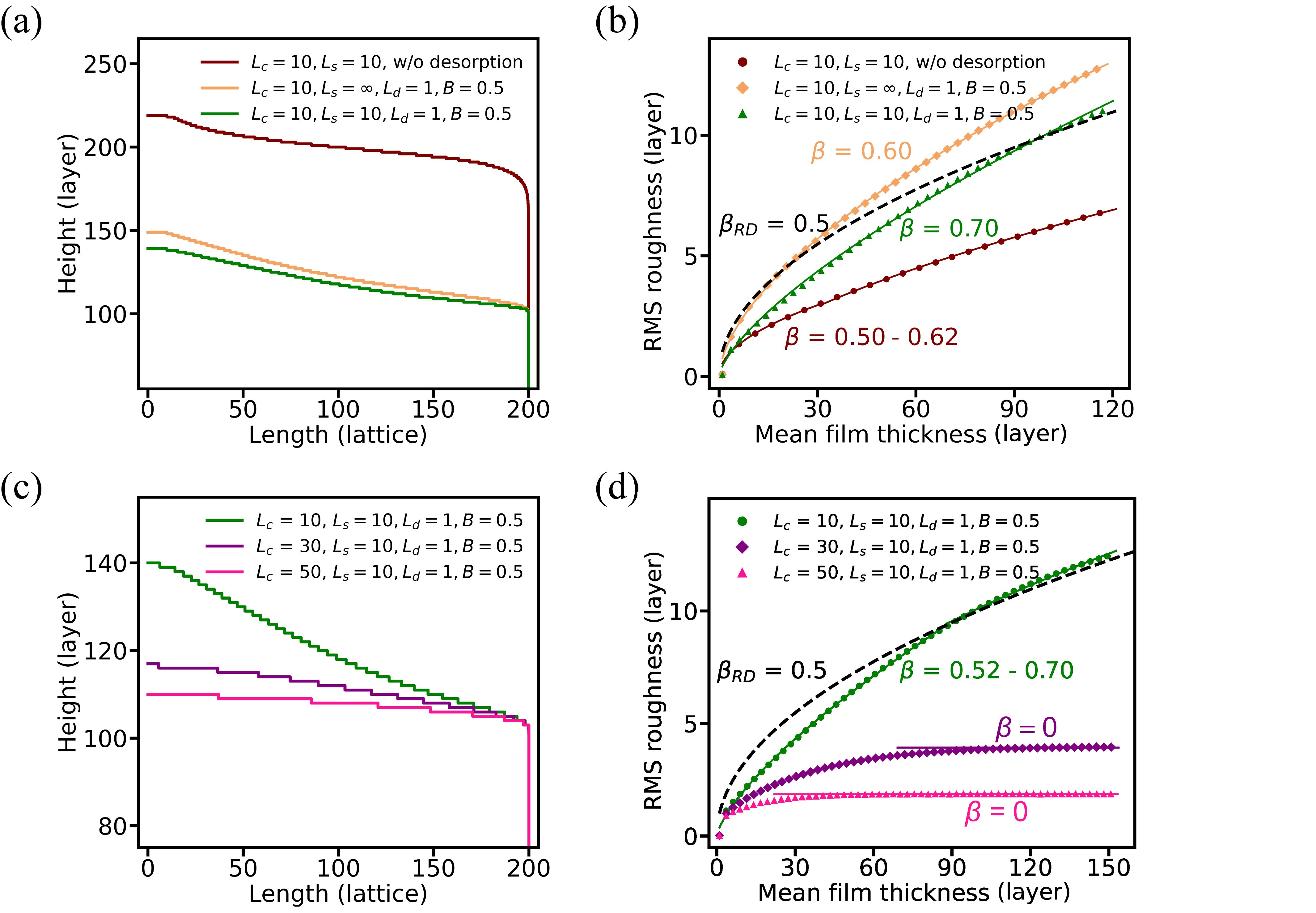}
\caption{Simulated mound formation. (a) Attempts to mimic the experimental mound shapes by varying the Schwoebel length and desorption parameters.  (b) RMS roughness obtained from the mound shapes.  (c, d) transition to flat-topped mounds as the critical radii for next-layer nucleation is increased.  $L_c$ of 10, 30, and 50 correspond to  $\Lambda/L_c$  = 20, 6.67, and 4.0 respectively.}
\label{Fig8.main}
\end{figure*}

The above results show that that the modified LSF model correctly predicts that step edges with larger terrace lengths ($L\gg L_s$) cover their own terraces more slowly than they would  with perfect sticking and incorporation. However, the model predicts highly asymmetric mound profiles that we did not observe in our film-growth experiment.  Rather, the experiments showed mounds with flattened tops. This behavior can be reproduced in the LSF model by increasing the critical radius $L_c$ for nucleation of a new top layer.  Fig. 8(a) shows a comparison of the results with $L_c$ = 10, and   Fig. 8(b) shows the root mean square (RMS) roughness vs. mean film thickness for the same parameters. It is important to note that with an infinite Schwoebel barrier ($L_s=\infty$) and significant desorption ($B$ = 0.5), the roughness exceeded the random deposition curve (dashed line)  from almost  the beginning of the deposition. This rapid roughening is associated with the asymmetric profile, as we have discussed, although with $L_c>0$ the asperity is truncated and the asymmetry reduced.  For cases with smaller $L_s$, roughening is suppressed; however, the exponents can still be larger than 0.5 even without desorption [lower curve in Fig 8(b)].  This reproduces a counter-intuitive observation from the experiment; that is, a large exponent is observed even in the presence of strong relaxation (small $L_s$).  We found that this effect was more pronounced when desorption was also active [middle curve in Fig 8(b)]. Overall, models that combine interlayer transport, desorption, and a critical radius resemble the experiments in at least three aspects: (i) the mound shape is nearly symmetric but truncated with a flat top; (ii) the RMS roughness is less than that predicted without interlayer relaxation; and (iii) $\beta>0.5$, although since unstable growth leads to continuously increasing slopes,  and thus decreasing terrace lengths, the growth mechanisms converge at late times towards the case of an infinite step-edge barrier without desorption.  Thus, we expected a gradual transition back to $\beta=0.5$ at late times. This effect is observed in the results: \textcolor{black} {from the bottom curve in Figure 8(b), $\beta$ decreases to $\sim0.5$ after roughly 60 layers while at the begining it is 0.62.}  In these cases the exponent decreases as the film thickness increases.

%where a range of exponents is given%

As $L_c$ was increased further, a  new regime was encountered.   Politi and Villain found that when the ratio of the mound radius $\Lambda$ to the critical radius $L_c$ becomes smaller than   $\Lambda/L_c = 9.397$ in 1+1 dimensions, the mounds become stable.\cite{Politi}   We have reproduced this effect in Fig. 8(c, d), which  shows a study varying $L_c$.  In the next section we discuss higher temperature growth where  flat featureless surfaces without obvious grooves or grain boundaries are observed, which may be a result of a morphological transition to stable mound growth.

\subsection*{In-situ XPCS results for 120$^\circ$C  growth}

\begin{figure*}[!ht]
\centering
\includegraphics[width=1\textwidth]{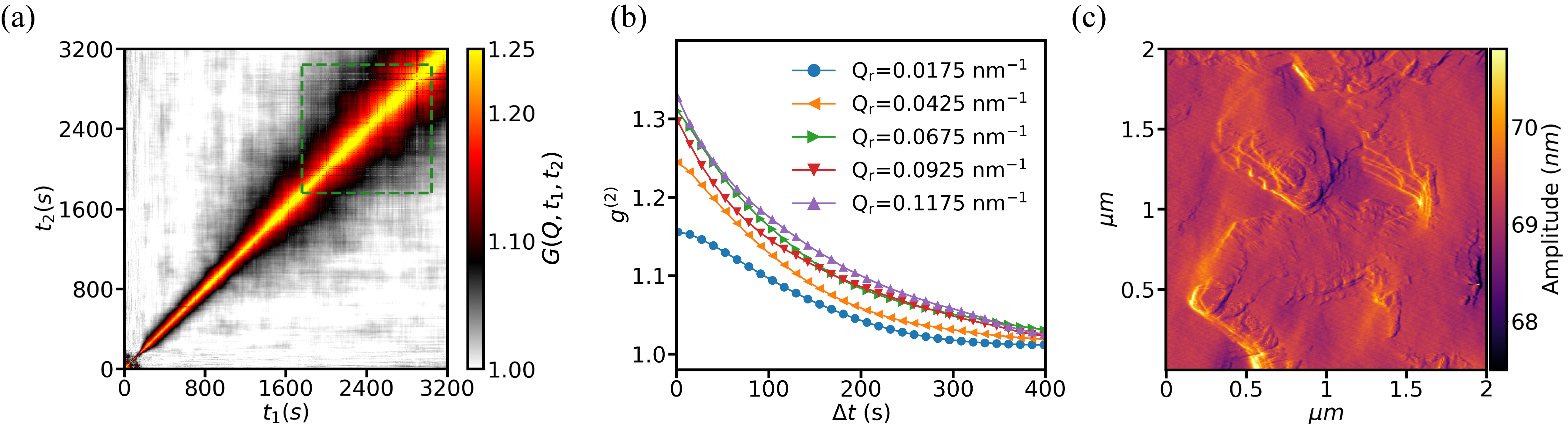}
\caption{DIP deposition on SiO$_2$ at 120 $^{\circ}$C. (a) Two-time autocorrelations plot of the complete deposition at 120 $^{\circ}$C, shutter opens at 40 s and closes at 3200 s ($Q_{r}$ = 0.0925 nm$^{-1}$; $Q_{z}$ = 0.4125 nm$^{-1}$). A zoomed in image to show the first layer growth can be found in Supplementary Fig. 3. (b) Autocorrelations averaged over 1760 s to 3040 s as a function of $Q_{r}$; $Q_{z}$ is fixed at 0.4125 nm$^{-1}$. (c) AFM image of a thin film growth in two layers with a temperature of 120 and 140 $^{\circ}$C for the first and second layers respectively, the mound boundaries become inapparent at these growth temperatures.}
\label{Fig9.main}
\end{figure*}

The growth mode at higher temperatures is different from that at lower temperatures, as shown in Fig. 9 and Supplementary Figs. 3. The first layer forms in a two-dimensional mode like the lower temperature growth; however,  there is no evidence from coherent X-ray results showing local step flow because there are no off-diagonal streaks in the two-time plot or oscillations in one-time correlations.  Fig. 9(a,b) shows the XPCS results for the later part of the  120$^\circ$C film growth, which confirms the absence of off-diagonal streaks in the two-time plot and the overall decay of the one-time correlations is somewhat faster than that in  the data for lower temperature growths.  Fig. 9(c) shows the post-deposition AFM scan, which shows that the surface is generally flat. Although there were a few high spots, no distinct mounds were observed.  Both the first monolayer and bulk sticking coefficients are reduced relative to lower temperature deposition (0.68 and 0.778, respectively in Table 1 and Supplementary Table 2).   In Fig. 6(b), we showed that desorption tends to flatten the edges of the mounds. Therefore, it is reasonable to infer that at 120$^{\circ}$C  the desorption effect promotes healing of the grooves between mounds. Because there are only several step edges on each mound the local step flow motion is no longer visible in the coherent X-ray measurements.

\section*{Discussion}

The principal results of this study on diindenoperylene (DIP) growth are as follows: (i) coherent X-ray scattering in the GISAXS geometry is sensitive to the spatial correlation of islands from one monolayer to the next during the transition from two-dimensional growth to mounded growth; (ii) step-flow dynamics exhibit a terrace-length dependent step velocity arising from the complex interplay of step-edge barrier and desorption effects; and (iii) at higher growth temperatures (120$^\circ$C) mounds merge into a relatively featureless film, possibly signaling a transition to stable mound growth. We briefly discuss each of these findings below.

\subsection*{Growth mode transition}
Our observation in Fig. 2 that only a single monolayer forms in a two-dimensional mode before the transition to mounded growth is surprising because previous investigations have suggested a gradual transition.\cite{durr2006, kowarik2008, frank2014}    For example, Kowarik et al. reported layer-by-layer growth for the first seven monolayers during DIP growth on SiO$_2$ at 130$^\circ$C.\cite{kowarik2008}    Our measurements for 80$^\circ$C growth in Fig. 2 suggest that the first monolayer reaches a coverage of $\gtrsim$90\% before the second layer nucleates. However, the diffuse scattering intensity profiles in Fig, 2(a, b) do not exhibit  intensity oscillations beyond the first layer, which suggest that the second and subsequent layers do not reach such high coverages before the nucleation of the next layer.   An exception in our data that more closely resembles the prevailing literature results is for growth at 40$^\circ$C (Supplementary Fig. 3) where we observed weak diffuse scattering intensity oscillations for three monolayers. There may be differences in experimental conditions, such as the deposition rate, that produce differences in the growth behavior.

 We found that the second and subsequent monolayer island growth positions are not correlated with the first layer.  This behavior gives rise to the distinctive two-time correlation plot shown in Fig. 2(c) for growth at 80$^\circ$C. The effect was more robust than the observation of layered growth, since it was consistent across all of the results over the range of  temperatures studied (see Supplementary Fig. 3). While there have been relatively few reports of coherent X-ray scattering measurements during epitaxial growth that we can compare to, a recent work by Ju et al.  showed that layer-by-layer growth of GaN produces a  ``checkerboard'' pattern in two-time correlation plots.\cite{Ju2019}     The results in Fig. 2(c) exhibit a similar behavior for the first monolayer of growth before the mode switches abruptly.  In particular, there is no strong correlation between the first monolayer growth (40-110 s) and the second monolayer (110-160 s), indicating that the second monolayer nucleates at positions that are uncorrelated with the first.  We speculate that  the positions of the second monolayer nuclei are unconstrained because the first monolayer closes up quickly.  The results of Ju et al. are instructive because they illustrate how the correlations of island positions from one layer to the next produces a checkerboard pattern in the two-time correlation function. Conversely, the absence of a checkerboard pattern in our data suggests that they are uncorrelated during the first few layers of growth. This result is also consistent with another (low coherence X-ray scattering) study of DIP growth on SiO$_2$  by Frank et al., who found an abrupt increase in the island size and spacing after the first monolayer.\cite{frank2014}  This behavior was associated with differences in surface diffusion energy barriers and island incorporation energies for DIP molecules on SiO$_2$ versus on DIP.

As we have already discussed, the situation changes once the mound shapes begin to stabilize at approximately 300 s in Fig 2(c) because each new layer is constrained to nucleate on the tops of the wedding-cakes and the entire stack of layers propagate radially out from nearly fixed center positions. This behavior gives rise to the off-diagonal streaks.  We also note that there was no modulation of the correlations along the streaks because nucleation at the tops of different mounds fell out of phase almost immediately.  Thus, the streaks were mainly due to the propagation of step edges, rather than being a direct result of nucleation events.

\subsection*{Step flow dynamics}
 It is known that large step edge barriers induce a mounded morphology by limiting the interlayer transport of ad molecules. However, a large roughening exponent has not previously been explained in this context. We find both experimental and computational evidence that roughening with  $\beta >$ 0.5 occurs, arising from step edge barriers and desorption.   This effect is evidently due to the terrace length dependence of both effects where unstable ($\beta>0$) growth results in increasing slopes and decreasing terrace lengths which boost the instabilities relative to the very early times. According to our modeling, the growth exponent should asymptotically converge to the random deposition value $\beta_{RD} =$0.5, because interlayer transport and desorption become negligible in the limit of short terrace-lengths. We conclude that rapid roughening may occur on  the timescale of some experiments, but the effect is  transient and $\beta$ does not exhibit a universal value distinct from random deposition.

Desorption is most noticeable in suppressing the growth rate, but it also influences the step motion and the mound profile by modifying the dispersion of the step velocity.  In coherent X-ray scattering the diffuse signal from the average mound profile, which normally acts as an obscuring background to the weaker signal from step-edge arrays, becomes a quasi-static reference that makes XPCS extremely  sensitive to the motion of step arrays.\cite{Headrick2019}  This mode is known as  \emph{heterodyne mixing}  which normally requires an external static reference field.    However, in certain cases, the static reference signal can originate from the sample.\cite{deJeu2005}  We have utilized this technique to identify a previously unobserved effect where the step velocity, or more accurately the time for a step to advance by the length of its lower terrace, is affected by terrace-length dependent desorption.  Although this is a plausible result, there are several additional questions that could be addressed. For example,  why do the coherent oscillations in Fig. 3 damp out quickly compared with the simulated results?  Considering this question, we note that in our model,  we have   assumed that nucleation always occurs in the center of the layer island below. The AFM images (i.e. Fig. 4) show that this is not necessarily the case, because many of the mounds have large differences in the slopes on opposite sides. We have also assumed that nucleation is timed to occur exactly when the underlying layer reaches the critical radius, whereas in a more complete model, it should be stochastic.  Another effect is that the overall mound shape is not stationary during unstable mound growth, so that it produces a ``quasi-static'' reference field, while a completely static reference would be ideal for characterizing the correlations with heterodyne mixing.  This effect may be dominant in the very early stages of mound formation when the correlation oscillations are not observed because the mound shape changes rapidly on the timescale of monolayer deposition.    

\subsection*{High temperature regime}
For growth at 120$^\circ$C and above we observed flat, featureless growth, and we have identified several mechanisms that may account for these observations. First, since desorption is inhibited where the slope is high, the grooves between mounds may tend to ``heal''.  Second,   a large critical radius  blunts the high-slope asperity at the tops of mounds,\cite{Krug2002} and it can lead to stable mound growth if the critical radius is sufficiently large.\cite{Politi}  Experimentally, the surface is stable in the sense that mounds and step arrays with slopes that  continually increase with growth time are not observed.   Within the range of temperatures studied, the films did not  sublimate significantly after deposition at the growth temperature.  We note that sublimation would likely result from the re-emission of monomers from step edges, a mechanism that we did not include in our modeling.  

\section*{Conclusions}
In conclusion, a combined \emph{in-situ} XPCS and \emph{ex-situ} X-ray and AFM studies was carried out to investigate the dynamic behavior during oriented polycrystalline  thin film growth of diindenoperylene (DIP). Nearly complete first layer growth was observed, followed by a transition to mounded growth. Detailed information about the local step motion  was obtained, and oscillations in the correlations arising from local step flow were observed.  A numerical model was constructed with terrace-length-dependent desorption, which  predicted behavior similar to the experiments. Desorption has long been neglected in constructing roughening models of surface growth, and our work suggests that it may be an important missing piece in the puzzle of surface dynamics and thin film morphology.

\section*{Acknowledgements}

The authors acknowledge the contributions of R. Greene for technical support at the CHX beamline; and to Nicole Bouffard of the Microscopy Imaging Center at the University of Vermont for technical assistance with the AFM imaging. This work was supported by the U. S. Department of Energy (DOE) Office of Science under Grant No. DE‐SC0017802. K.F.L. and P.M. were supported by the National Science Foundation (NSF) under grant no. DMR-1709380. This research used the 11-ID and 4-ID beamlines of the National Synchrotron Light Source II, a U.S. DOE Office of Science User Facility operated for the DOE Office of Science by the Brookhaven National Laboratory under Contract No. DE-SC0012704.

\vspace*{1\baselineskip} 
\bibliographystyle{unsrt}
\bibliography{ref}

\end{document}